\documentclass[aps,preprint,showpacs,preprintnumbers,amsmath,amssymb]{revtex4}
\usepackage{amsmath,mathrsfs,amsbsy,color,graphicx,bm,amsthm,amsfonts}
\usepackage{units}
\usepackage{bbm}
\usepackage{times}
\usepackage{dcolumn}
\usepackage{mathrsfs}
\usepackage{amsmath,amssymb,epsfig}
\begin{document}

\title{ Entangled Unruh-DeWitt detectors amplify quantum coherence }
\author{Shu-Min Wu$^1$\footnote{Email: smwu@lnnu.edu.cn}, Yu-Xuan Wang$^1$, Wentao Liu$^2$\footnote{wentaoliu@hunnu.edu.cn (corresponding authors)} }
\affiliation{ $^1$ Department of Physics, Liaoning Normal University, Dalian 116029, China\\
$^2$ Department of Physics, Hunan Normal University, Changsha 410081, China}


\begin{abstract}
We explore the quantum coherence between a pair of entangled Unruh-DeWitt detectors, interacting with a quantum field, using a nonperturbative approach in a (3+1)-dimensional Minkowski spacetime with instantaneous switching ($\delta$-switching). It is intriguing to observe that for a maximally entangled state, increasing the coupling strength enhances the detectors' initial quantum coherence while simultaneously causing a monotonic decrease in their initial entanglement. This reveals a remarkable phenomenon: through nonperturbative interactions, entangled Unruh-DeWitt detectors can exhibit a dual effect-amplifying quantum coherence while degrading quantum entanglement. This finding stands in stark contrast to previous studies based on perturbative methods or Gaussian switching functions, which generally concluded that interactions between detectors and the field lead to a simultaneous degradation of quantum coherence and entanglement due to environmental decoherence. Notably, while initially separable detectors successfully harvest quantum coherence from the vacuum, entanglement extraction remains fundamentally prohibited. These contrasting behaviors underscore the fundamental distinction between coherence and entanglement as quantum resources, and highlight their complementary roles in field-detector interactions.
\end{abstract}

\vspace*{0.5cm}
 \pacs{04.70.Dy, 03.65.Ud,04.62.+v }
\maketitle
\section{Introduction}
Quantum coherence, arising from the superposition of quantum states, represents a fundamental feature of quantum physics \cite{L1}. Its generation and preservation are essential for quantum information processing tasks \cite{L2}. Although coherence is central to quantum theory, it only began receiving widespread attention after Baumgratz $et$ $al.$ introduced an information-theoretic framework for its quantification, proposing the \( l_1 \)-norm and the relative entropy as coherence measures \cite{L47}. Similar to quantum entanglement, quantum coherence is a crucial quantum resource, playing key roles in diverse fields such as quantum networks, quantum teleportation, quantum batteries, quantum materials,  quantum computation, and quantum biology \cite{L4,L5,L6,L7,L8,L9,L10,L11,L12,L13,L13-1}. It captures the essence of quantum mechanics in systems of arbitrary dimension, from single-qubit superposition to quantum correlations in multipartite systems \cite{L2,N1}. Notably, quantum entanglement stems from nonlocal superpositions and has been shown to be obtainable from quantum coherence without introducing additional coherence. This suggests that an entanglement monotone can induce a corresponding coherence monotone for a given quantum state \cite{L14,W1}. While significant progress has been made in understanding the relationship between coherence and entanglement, it remains a rich topic for ongoing research.

The Minkowski vacuum, as a striking example of nonclassical correlations, continues to be a compelling subject in quantum field theory and gravitational physics, with many of its properties still not fully understood \cite{E26,E27,E28}. In free quantum field theory on Minkowski spacetime, the vacuum state exhibits strong entanglement across spacelike-separated regions \cite{L15}. Building on this foundational insight, Summers and Werner, using algebraic methods, demonstrated that field observables in such regions can produce maximal violations of Bell inequalities \cite{L17,L18,L19}, underscoring the deeply nonclassical nature of vacuum entanglement. To explore this phenomenon more concretely, researchers have developed models involving localized particle detectors coupled to quantum fields-most notably, the Unruh-DeWitt (UDW) detector model \cite{L23,L24,L25}. In this framework, a detector is treated as a localized quantum system (or a particle in a ``box") that interacts with the field only when the ``box" is opened. Remarkably, when multiple such detectors are individually coupled to the vacuum without any direct interaction between them, they can nonetheless become quantum mechanically entangled. This reveals that the vacuum itself can serve as a physically accessible reservoir of entanglement.
This phenomenon is known as quantum entanglement harvesting-the extraction of entanglement from quantum fields via localized interactions \cite{L20,L21,L22}.  Over time, entanglement harvesting has evolved into a powerful theoretical tool for probing how nontrivial features of spacetime influence vacuum correlations. These include effects from spacetime curvature, nontrivial topology, the Unruh and Hawking effects, and various cosmological scenarios \cite{L36,L37,L38,L39,L40,L41,L42,L42-3,L42-4,L42-5,L42-7,L42-8,L42-9,L42-10,L42-11,L42-12,L42-13,L42-15,L26,L27,L28,L29,L30,L31,L32}.

When two UDW detectors interact with a quantum field, the quantum entanglement of an initially maximally entangled state typically undergoes degradation. This degradation is primarily attributed to the monogamy of entanglement, a constraint rigorously formulated by Coffman, Kundu, and Wootters \cite{ckw}. In contrast, quantum coherence is not subject to monogamy, making it operationally distinct from entanglement. This fundamental difference naturally leads to an important question: can quantum coherence be enhanced beyond its initial value during nonperturbative interactions between detectors and the field, even as entanglement degrades? This question not only challenges conventional perspectives on the dynamics of quantum resources in the UDW detector model, but also forms a central motivation of our work.
Moreover, previous studies have shown that detectors prepared in a separable initial state  using a nonperturbative approach generally fail to harvest quantum entanglement from the vacuum \cite{E}. Yet, since quantum coherence is widely regarded as a necessary precursor to entanglement, it raises another key question: can initially separable detectors harvest quantum coherence from the vacuum solely through their interaction with the field?
Together, these questions drive our investigation into the fundamental differences between coherence and entanglement in quantum field settings and offer new insights into the resource dynamics of the UDW detector framework.

To this end, we investigate quantum coherence through a nonperturbative analysis of two inertial UDW detectors interacting with a scalar field in (3+1)-dimensional Minkowski spacetime. The detectors undergo instantaneous switching ($\delta$-switching) and are coupled to the field with full consideration of communication channels between them. Our results reveal two key features: (a) when the detectors are initially prepared in a maximally entangled state, increasing the coupling strength results in a pronounced amplification of quantum coherence, while entanglement degrades monotonically;  (b) when the detectors start in a separable state, entanglement harvesting remains forbidden, yet a nontrivial amount of quantum coherence can still be extracted from the vacuum.
These findings stand in stark contrast to earlier results from perturbative or Gaussian-switching-based studies, which typically found that field-detector interactions lead to simultaneous degradation of both coherence and entanglement due to environmental decoherence \cite{rbm1,rbm2,rbm3}. In contrast, our work highlights that quantum coherence and entanglement play fundamentally different roles in field-detector interactions, especially in nonperturbative regimes, where coherence emerges as a more robust and accessible quantum resource than entanglement.

The structure of this paper is organized as follows. In Section II,  we briefly review the UDW detector model with \(\delta\)-switching and present the reduced density matrix of the detectors after their interaction with the quantum field.
In Section III,  we analyze the quantum coherence between a pair of initially entangled UDW detectors interacting with a quantum field, employing a nonperturbative framework in  Minkowski spacetime. Finally, Section IV summarizes our main results and conclusions.
Throughout the paper, we adopt natural units (\(\hbar = c = 1\)), assume that the detectors  denote spacetime points as \(X = (t, \mathbf{x})\).

\section{UDW DETECTORS WITH DELTA SWITCHING}
\subsection{Time-evolution operator}
Consider two UDW detectors, $A$ and $B$, each locally coupled to a quantum scalar field  $\widehat{\phi}(X)$.
The interaction Hamiltonian for a linearly coupled detector \(j\)$\in\{A,B\}$ can be expressed as
\begin{eqnarray}\label{1}
\widehat{H}^{\tau_{j}}_{j}(\tau_{j})=\lambda_{j}\chi_{j}(\tau_{j})\widehat{\mu}
_{j}(\tau_{j})\otimes\int \textrm{d}^{n}xF_{j}(\textbf{x}-\textbf{x}_{j})\widehat{\phi}(X_{j}(\tau_{j})),
\end{eqnarray}
 where, $\tau_{j}$ is the detector's proper time, $\lambda_{j}$ represents the coupling constant, $\chi_{j}(\tau_{j})$
denotes the switching function, and $F_{j}(\textbf{x}-\textbf{x}_{j})$ is the smearing function of detector \(j\).
The superscript $\tau_{j}$ on the Hamiltonian indicates that it generates time translations with respect to
$\tau_{j}$. The switching function $\chi_{j}(\tau_{j})$ governs the temporal profile of the coupling, while the smearing function  $F_{j}(\textbf{x}-\textbf{x}_{j})$ determines the detector's spatial extent. The operator $\widehat{\mu}_{j}(\tau_{j})$, corresponding to the monopole moment of detector, is defined as
\begin{eqnarray}\label{2}
\widehat{\mu}_{j}(\tau_{j})=|e_{j}\rangle\langle g_{j}|e^{i\Omega_{j}\tau_{j}}+|g_{j}\rangle\langle e_{j}|e^{-i\Omega_{j}\tau_{j}},
\end{eqnarray}
where $|g_{j}\rangle$ and $|e_{j}\rangle$ denote the ground and excited states of the detector, respectively, and \(\Omega_j\) is the energy gap between them. For our analysis, we adopt an instantaneous ($\delta$-function) switching scheme described by $\chi_{j}(\tau_{j})=\eta_{j}\delta(\tau_{j}-\tau_{j,0})$ \cite{E11}, where  \(\tau_{j,0}\) is the proper time at which detector \(j\) interacts with the field, and $\eta_{j}=\int^{\infty}_{-\infty}\chi_{j}(\tau_{j})\textrm{d}\tau_{j}$ is a constant with dimensions of time.

By employing $\delta$-function switching, we analyze the quantum coherence between two detectors in a nonperturbative framework. We adopt a Cartesian coordinate system \((t, \textbf{x})\) in a (3+1)-dimensional Minkowski spacetime to conveniently specify the positions of the detectors. Since the coordinate time \( t \) can be used to label events for both detectors, their respective proper times
\( \tau_{j} \) can be expressed as functions of  \( t \), i.e., \( \tau_{j}(t) \). With this shared time parameter, the time-evolution operator in the interaction picture takes the form $\widehat{U}_{I}=\mathcal{T}_{t}\, \textrm{exp}[-\textrm{i}\int_{\Re}\textrm{d}t\widehat{H}^{t}_{I}(t)]$,
where \( \mathcal{T}_{t} \) denotes the time-ordering operator with respect to \( t \), and the Hamiltonian \( \widehat{H}_{I}^{t}(t) \) is given by $\widehat{H}_{I}^{t}(t)=\frac{\textrm{d}\tau_{A}}{\textrm{\textrm{d}}t} \widehat{H}_{A}^{\tau_{A}}(\tau_{A}(t))+
\frac{\textrm{d}\tau_{B}}{\textrm{d}t} \widehat{H}_{B}^{\tau_{B}}(\tau_{B}(t))$ \cite{E38,E39}.
Assuming detector $A$ is switched on before detector $B$, i.e., \( t(\tau_{A,0}) \leq t(\tau_{B,0}) \), the delta switching allows us to write the time-evolution operator as
\begin{eqnarray}\label{3}
\widehat{U}_{I}=\textrm{exp}[\widehat{\mu}_{B}(\tau_{B,0}) \otimes \widehat{Y}_{B}(\tau_{B,0})]\textrm{exp}[\widehat{\mu}_{A}(\tau_{A,0}) \otimes \widehat{Y}_{A}(\tau_{A,0})],
\end{eqnarray}
with $\widehat{Y}_{j}(\tau_{j,0}):=-\textrm{i}\lambda_{j}\eta_{j}\frac{\textrm{d}\tau_{j}}{\textrm{d}t}\mid_{\tau_{j,0}}\int \textrm{d}^{n}xF_{j}(\mathbf{x-x}_{j})
\widehat{\phi}(X_{j}(\tau_{j,0})).$
The operator \( \widehat{Y}_{j} \) can be interpreted as a smeared field operator evaluated at the switching time \( t(\tau_{j,0}) \) \cite{E11}. To quantify the coherence between the detectors, we define the commutator
\(\kappa\) and anti-commutator \(\omega\) as
\begin{eqnarray}\label{4}
\kappa:=-\textrm{i}\langle0|[\widehat{Y}_{A},\widehat{Y}_{B}]|0\rangle,
\end{eqnarray}
\begin{eqnarray}\label{5}
\omega:=2\langle0|\{{\widehat{Y}_{A},\widehat{Y}_{B}\}|0\rangle}.
\end{eqnarray}
Here, $\kappa$ corresponds to the Pauli-Jordan function and reflects the causal structure of the interaction: it is nonzero when the detectors are timelike or lightlike separated (i.e., communication is possible), and vanishes under spacelike separation. Conversely, $\omega$, related to the Hadamard distribution, can remain nonzero even for spacelike separation, capturing vacuum-induced quantum coherence between causally disconnected detectors.

\subsection{Density matrix}
We begin by considering a Bell-type entangled state of two detectors, given by
\begin{eqnarray}\label{6}
|\psi\rangle=\cos \theta|g_{A}g_{B}\rangle+\sin \theta|e_{A}e_{B}\rangle,
\end{eqnarray}
with $\theta\in[0,\frac{\pi}{2}]$. The state is separable when  $\theta=0$ or $\frac{\pi}{2}$, and entangled otherwise. In particular, the detectors are maximally entangled when  $\theta=\frac{\pi}{4}$.
Assuming the quantum field is in the vacuum state \( |0\rangle \), and  we analyze the detectors in the computational basis $\{|g_{A}g_{B}\rangle\ , |g_{A}e_{B}\rangle\ , |e_{A}g_{B}\rangle\ ,|e_{A}e_{B}\rangle\}$.  After interaction with the field, the reduced density matrix  \( \rho_{AB} \) of the two detectors is given by
\begin{eqnarray}\label{7}
\rho_{AB}=\textrm{Tr}_{\phi}[\widehat{U}_{I}(\rho_{AB,0}\otimes|0\rangle\langle0|)\widehat{U}^{\dag}_{I}],
\end{eqnarray}
\begin{eqnarray}\label{8}
\scalebox{1}{$=
 \left(\!\!\begin{array}{cccccccc}
\rho_{11} & 0 & 0 & \rho_{14}\\
0 & \rho_{22} & \rho_{23} & 0\\
0 & \rho_{23}^{*} & \rho_{33} & 0\\
\rho_{14}^{*} & 0 & 0 & \rho_{44}
\end{array}\!\!\right)
$},
\end{eqnarray}
where $\rho_{AB,0}=|\psi\rangle\langle\psi|$. We further assume that the two detectors are inertial and have no relative motion (i.e., \( \frac{\textrm{d}\tau_{j}}{\textrm{d}t} = 1 \)) in  Minkowski spacetime.  They interact with a massless, minimally coupled scalar field, which is initially in the vacuum state. Under these conditions, the explicit expressions for the elements of the reduced density matrix
 \( \rho_{AB} \)  of the two detectors can be derived as
\begin{eqnarray}\label{9}
\rho_{11}&=&\frac{1}{4}[1+f_{A}f_{B}\cosh\omega+(2\cos^{2}\theta-1)(f_{A}+f_{B}\cos(2\kappa))]\notag\\
&+&\frac{1}{2}\cos\theta \sin\theta f_{B}[f_{A}\sinh\omega \cos\gamma-\sin(2\kappa)\sin\gamma]\notag,
\end{eqnarray}
\begin{eqnarray}\label{10}
\rho_{22}&=&\frac{1}{4}[1-f_{A}f_{B}\cosh\omega+(2\cos^{2}\theta-1)(f_{A}-f_{B}\cos(2\kappa))]\notag\\
&-&\frac{1}{2}\cos\theta \sin\theta f_{B}[f_{A}\sinh\omega \cos\gamma-\sin(2\kappa)\sin\gamma],\notag
\end{eqnarray}
\begin{eqnarray}\label{11}
\rho_{33}&=&\frac{1}{4}[1-f_{A}f_{B}\cosh\omega-(2\cos^{2}\theta-1)(f_{A}-f_{B}\cos(2\kappa))]\notag\\
&-&\frac{1}{2}\cos\theta \sin\theta f_{B}[f_{A}\sinh\omega \cos\gamma+\sin(2\kappa)\sin\gamma],\notag
\end{eqnarray}
\begin{eqnarray}\label{12}
\rho_{44}&=&\frac{1}{4}[1+f_{A}f_{B}\cosh\omega-(2\cos^{2}\theta-1)(f_{A}+f_{B}\cos(2\kappa))]\notag\\
&+&\frac{1}{2}\cos\theta \sin\theta f_{B}[f_{A}\sinh\omega \cos\gamma+\sin(2\kappa)\sin\gamma],\notag
\end{eqnarray}
\begin{eqnarray}\label{13}
&\rho_{14}&e^{\textrm{i}(\Omega_{A}\tau_{A,0}+\Omega_{B}\tau_{B,0})}=\frac{f_{B}}{4}[f_{A}\sinh\omega+\textrm{i}(2\cos^{2}\theta-1)\sin(2\kappa)]\notag\\
&+&\frac{1}{2}\cos\theta \sin\theta[(1+f_{A}f_{B}\cosh\omega)\cos\gamma
+\textrm{i}(f_{A}+f_{B}\cos(2\kappa))\sin\gamma],\notag
\end{eqnarray}
\begin{eqnarray}\label{14}
&\rho_{23}&e^{\textrm{i}(\Omega_{A}\tau_{A,0}+\Omega_{B}\tau_{B,0})}=-\frac{f_{B}}{4}[f_{A}\sinh\omega+\textrm{i}(2\cos^{2}\theta-1)\sin(2\kappa)]\notag\\
&+&\frac{1}{2}\cos\theta \sin\theta[(1-f_{A}f_{B}\cosh\omega)\cos\gamma
+\textrm{i}(f_{A}-f_{B}\cos(2\kappa))\sin\gamma],\notag
\end{eqnarray}
with
\begin{eqnarray}\label{15}
\gamma:=\Omega_{A}\tau_{A,0}+\Omega_{B}\tau_{B,0} ,
\end{eqnarray}
\begin{eqnarray}\label{16}
\alpha_{j}(\textbf{k}):=-\textrm{i}\frac{2\lambda_{j}\eta_{j}}{\sqrt{2|\textbf{k}|}}\widetilde{F}_{j}^{*}(\textbf{k})e^{\textrm{i}|\textbf{k}|
\tau_{j,0}-\textrm{i}\textbf{k}\cdot \textbf{x}_{j}},
\end{eqnarray}
\begin{eqnarray}\label{17}
f_{j}:=\textrm{exp}(-\frac{1}{2}\int \textrm{d}^{n}k|\alpha_{j}(\textbf{k})|^{2})\in\Re,
\end{eqnarray}
\begin{eqnarray}\label{18}
\kappa=\frac{\textrm{i}}{4}\int \textrm{d}^{n}\textbf{k}[\alpha^{*}_{A}(\textbf{k})\alpha_{B}(\textbf{k})-\alpha_{A}(\textbf{k})\alpha^{*}_{B}(\textbf{k})]\in\Re,
\end{eqnarray}
\begin{eqnarray}\label{19}
\omega=-\frac{1}{2}\int \textrm{d}^{n}\textbf{k}[\alpha^{*}_{A}(\textbf{k})\alpha_{B}(\textbf{k})+\alpha_{A}(\textbf{k})\alpha^{*}_{B}(\textbf{k})]\in\Re,
\end{eqnarray}
where $\widetilde{F}_{j}(\textbf{k})$ denotes the Fourier transform of the smearing function. Notably, the explicit form of the density matrix allows for a straightforward and systematic analysis of the correlation dynamics between the detectors. In particular, several matrix elements vanish under specific choices of the parameter $\theta$, such as $\theta=0$, $\theta=\frac{\pi}{4}$, $\theta=\frac{\pi}{2}$, significantly simplifying the overall structure of $\rho_{AB}$. These special cases are especially useful for exploring in greater detail how the initial state parameters influence the evolution of quantum correlations, including entanglement and coherence, between the two detectors.

Finally, we specify the spatial profile of the detectors by choosing a Gaussian smearing function, given by
\begin{eqnarray}\label{20}
F(\textbf{x})=\dfrac{1}{ (\sqrt{\pi} \sigma )^n } e^{ -\bm{x}^2/\sigma^2 }, \label{eq:Gaussian smearing}
\end{eqnarray}
where \( \sigma \)  characterizes the spatial width of the detector's smearing. With this choice, the key quantities introduced above can be expressed in more compact analytical forms as
\begin{eqnarray}\label{21}
f_{j}=\textrm{exp}(-\frac{\lambda^{2}_{j}\eta^{2}_{j}}{2\pi^{2}\sigma^{2}}),
\end{eqnarray}
\begin{eqnarray}\label{22}
\kappa=\frac{\lambda_{A}\lambda_{B}\eta_{A}\eta_{B}}{4\pi^{2}L\sigma}\sqrt\frac{\pi}{2}(e^{-(\Delta\tau+L)^2/2\sigma^{2}}
-e^{-(\Delta\tau-L)^2/2\sigma^{2}}),
\end{eqnarray}
\begin{eqnarray}\label{23}
\omega=-\frac{\lambda_{A}\lambda_{B}\eta_{A}\eta_{B}}{\sqrt{2}\pi^{2}L\sigma} \bigg[D^{+}(\frac{\Delta\tau+L}{\sqrt{2}\sigma})
-D^{+}(\frac{\Delta\tau-L}{\sqrt{2}\sigma})\bigg],
\end{eqnarray}
where $\Delta\tau:=\tau_{B,0}-\tau_{A,0}$, $L:=|\textbf{x}_{B}-\textbf{x}_{A}|$, and $D^{+}(x)$ is the Dawson function, defined by $D^{+}(x):=\frac{\sqrt{\pi}}{2}e^{-x^{2}}\mathrm{erfi}(x)$.

\section{Quantum coherence between a pair of entangled UDW detectors}
Quantum coherence, which stems from the principle of superposition, is a fundamental feature of quantum mechanics and serves as a necessary resource for the emergence of quantum entanglement. In this work, we investigate how the quantum coherence between a pair of entangled UDW detectors is affected by their nonperturbative interactions with a quantum field. We further compare the behavior of coherence with that of quantum entanglement to uncover their distinct and overlapping characteristics. To quantify coherence, we adopt two well-established measures: the  \( l_1 \)-norm of coherence and the relative entropy of coherence (REC)  \cite{L47}. For an \(n\)-dimensional quantum system described by a density matrix $\rho$, and with respect to a fixed reference basis \(\{ |i\rangle \}_{i=1,...,n}\), the $l_1$-norm of coherence is defined as the sum of the absolute values of all off-diagonal elements of
\(\rho_{AB}\)
\begin{eqnarray}\label{24}
C_{l_{1}}(\rho_{AB})=\sum_{i\neq j}|\rho_{i,j}|.
\end{eqnarray}
The REC, on the other hand, is given by the difference between the von Neumann entropy of the diagonalized state  \( S(\rho _{\textrm{diag}}) \)  and the entropy of the full state
\(\rho_{AB}\)
\begin{eqnarray}\label{25}
C_{REC}(\rho_{AB}) = S(\rho_{AB \textrm{diag}})-S(\rho_{AB}),
\end{eqnarray}
where \( \rho_{AB} \)  is the von Neumann entropy of the state \( \rho_{AB} \), and  \( S(\rho _{AB{\textrm{diag}}}) \)  is obtained by retaining only the diagonal elements of
\( \rho_{AB} \) in the reference basis \cite{L47}.

Employing Eqs.(\ref{8}) and  (\ref{24}), the \( l_1 \)-norm of quantum coherence for the bipartite detector state \( \rho_{AB} \) is obtained as
\begin{eqnarray}\label{26}
C_{l_{1}}(\rho_{AB})=2|\rho_{14}|+2|\rho_{23}|.
\end{eqnarray}
Next, we evaluate the REC induced by the quantum vacuum. To do so, we calculate the eigenvalues of the density matrix \( \rho_{AB} \). The structure of \( \rho_{AB} \) in Eq.(\ref{8}) yields four nonzero eigenvalues, given by
\begin{eqnarray}\label{27}
\lambda_{1}=\frac{1}{2}(\rho_{22}+\rho_{33}-\sqrt{\rho_{22}^{2} +4\rho_{23}^{2}-2\rho_{22}\rho_{33}+\rho_{33}^{2}}),
\end{eqnarray}
\begin{eqnarray}\label{28}
\lambda_{2}=\frac{1}{2}(\rho_{22}+\rho_{33}+\sqrt{\rho_{22}^{2} +4\rho_{23}^{2}-2\rho_{22}\rho_{33}+\rho_{33}^{2}}),
\end{eqnarray}
\begin{eqnarray}\label{29}
\lambda_{3}=\frac{1}{2}(\rho_{11}+\rho_{44}-\sqrt{\rho_{11}^{2} +4\rho_{14}^{2}-2\rho_{11}\rho_{44}+\rho_{44}^{2}}),
\end{eqnarray}
\begin{eqnarray}\label{30}
\lambda_{4}=\frac{1}{2}(\rho_{11}+\rho_{44}+\sqrt{\rho_{11}^{2} +4\rho_{14}^{2}-2\rho_{11}\rho_{44}+\rho_{44}^{2}}),
\end{eqnarray}
Accordingly, the REC for the state $\rho_{AB}$ takes the form
\begin{eqnarray}\label{31}
C_{REC}(\rho_{AB})=-\sum_{i}\zeta_{i}log_{2}\zeta_{i}+\sum_{j}\lambda_{j}log_{2}\lambda_{j},
\end{eqnarray}
where \( \zeta_{i} \) denote the diagonal elements of \( \rho_{AB} \) and \( \lambda_{j} \) are the nonzero eigenvalues of the full density matrix \( \rho_{AB} \).
To compare nonlocal coherence with quantum entanglement, we adopt negativity as a standard measure of bipartite entanglement. The entanglement between subsystems $A$ and $B$ in the joint state
is quantified by the negativity $N({\rho _{AB}})$, defined as
\begin{eqnarray}\label{32}
N({\rho _{AB}})=\frac{\|\rho _{(AB)}^{T_{A}}\|-1}{2},
\end{eqnarray}
where \( T_{A} \) represents the partial transpose of \( \rho _{AB} \) with respect to subsystem \( A \), and the symbol \( \|\cdot\| \) denotes the trace norm of a matrix \cite{L48}.  The quantity \( \|\rho^{T_A}_{AB}\| - 1 \) equals twice the sum of the absolute values of the negative eigenvalues of \( \rho^{T_A}_{AB} \). Using Eqs.(\ref{8}) and (\ref{32}), the negativity can be expressed in a compact and computable form as
\begin{eqnarray}\label{34}
N({\rho _{AB}})=\textrm{max}\bigg[0,\sqrt{|\rho_{14}|^{2}+(\frac{\rho_{33}-\rho_{22}}{2})^{2}}-
\frac{1}{2}(\rho_{22}+\rho_{33})\bigg].
\end{eqnarray}

\begin{figure}[htbp]
\centering
\includegraphics[height=1.8in,width=2.0in]{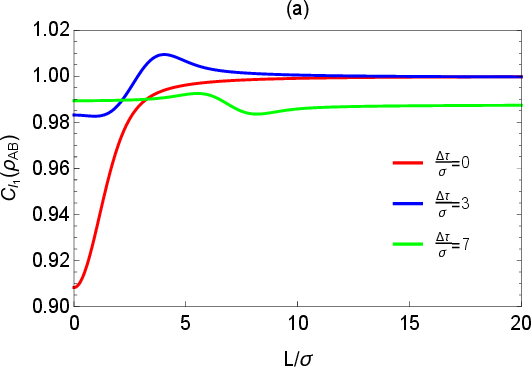}
\includegraphics[height=1.8in,width=2.0in]{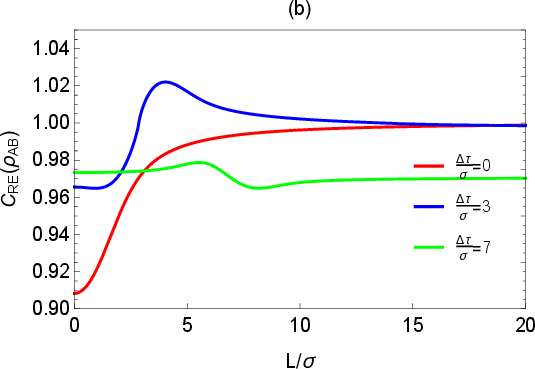}
\includegraphics[height=1.8in,width=2.0in]{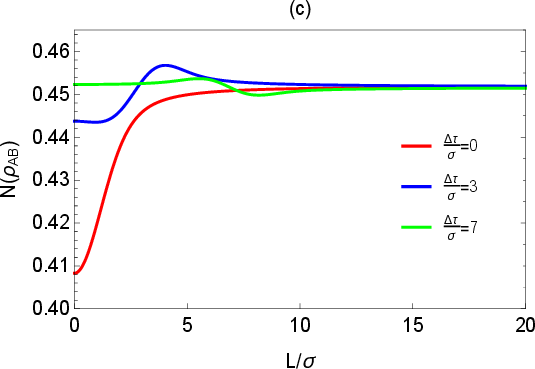}
\caption{Quantum coherence and entanglement between two UDW detectors as a function of the dimensionless detector separation $L/\sigma$, for various values of $\Delta\tau/\sigma$. The initial state parameter is fixed at $\theta=\pi/4$, and  detector parameters are set as $\lambda=\eta/\sigma=\Omega_{A}\sigma=\Omega_{B}\sigma=1$. }
\label{F1}
\end{figure}

In the following analysis, we explore the behavior of quantum coherence and entanglement using a spacetime diagram, with all quantities expressed in units of the characteristic  smearing width $\sigma$.  Fig.\ref{F1} shows the evolution of $C_{l_{1}}(\rho_{AB})$,  $C_{REC}(\rho_{AB})$, and $N({\rho _{AB}})$  for detectors initially prepared in a maximally entangled state
(i.e., \(\theta = \frac{\pi}{4}\)). These quantities are plotted as a function of the detector separation $L/\sigma$ for different values of \(\Delta \tau/\sigma\). For a maximally entangled initial state, the entanglement is initially  $N({\rho _{AB}})=1/2$, while the coherence measures are $C_{l_{1}}(\rho_{AB})=C_{REC}(\rho_{AB})=1$. It is immediately evident that entanglement is degraded under all circumstances, with the most significant degradation occurring at small detector separation $L/\sigma$. Interestingly, in this same regime, the quantum coherence can exceed its initial value. This suggests that even starting from a maximally entangled state, increasing the detector separation allows the system to extract additional coherence from the vacuum. This highlights a striking phenomenon: compared to the initial entanglement and coherence, while the separation suppresses entanglement, it can enhance coherence. The root of this contrast lies in their fundamental  differences-entanglement is restricted by monogamy, whereas coherence is not subject to such constraints. Moreover, when the detectors are nearly lightlike separated, particularly at  \(L = \Delta \tau\),
both coherence and entanglement exhibit noticeable oscillatory behavior due to the exchange of quanta. As the separation becomes large, these oscillations fade, and all quantities tend to stabilize, approaching constant asymptotic values.

\begin{figure}[htbp]
\centering
\includegraphics[height=1.8in,width=2.0in]{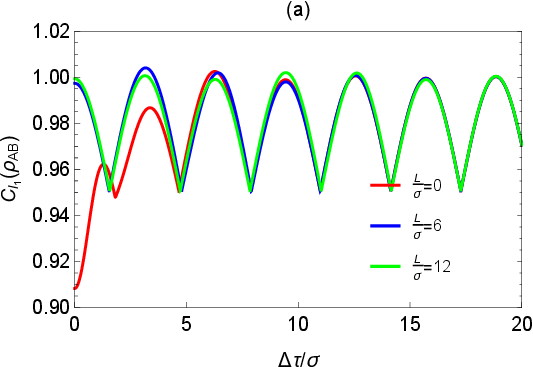}
\includegraphics[height=1.8in,width=2.0in]{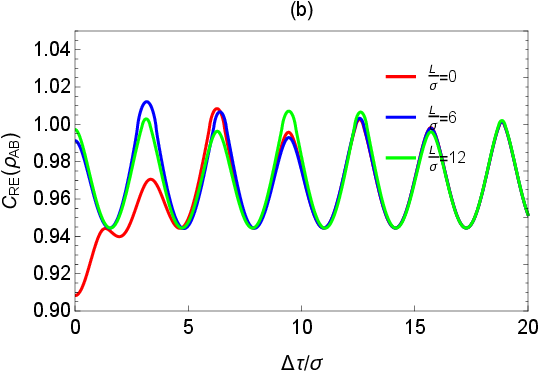}
\includegraphics[height=1.8in,width=2.0in]{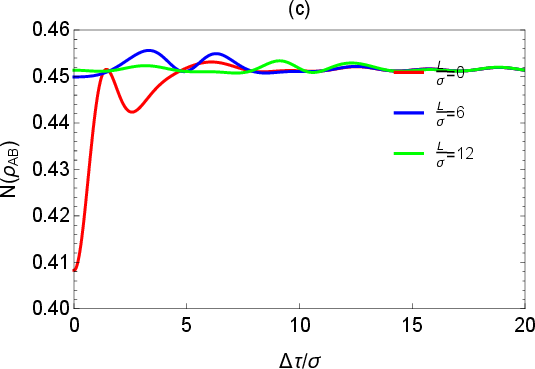}
\caption{Quantum coherence and entanglement  as a function of the switching difference $\Delta\tau/\sigma$ for various values of $L/\sigma$ with fixed $\theta=\pi/4$, $\lambda=\eta/\sigma=\Omega_{A}\sigma=\Omega_{B}\sigma=1$. }
\label{F2}
\end{figure}

In Fig.\ref{F2}, we present the behavior of quantum coherence and entanglement as a function of the switching time difference \(\Delta\tau/\sigma\) for various values of $L/\sigma$.
As illustrated, both quantum coherence and entanglement exhibit periodic variations with respect to
\(\Delta\tau/\sigma\), which originate from the phase factor $\gamma$ in the field correlations.
Notably, in the region where lightlike communication between the detectors is possible, both coherence and entanglement display pronounced distortions. This behavior highlights the sensitivity of quantum resources to causal quantum communication effects mediated by the field.
Furthermore, for an initially maximally entangled state, quantum coherence can, at certain values of \(\Delta\tau/\sigma\), exceed its initial value. In contrast, quantum entanglement remains consistently lower than its initial value across all switching differences. This contrast once again underscores the fact that quantum coherence, unlike entanglement, is not restricted by monogamy and can be more easily enhanced by vacuum fluctuations and interaction timing.

\begin{figure}
\begin{minipage}[t]{0.5\linewidth}
\centering
\includegraphics[width=3.0in,height=5.2cm]{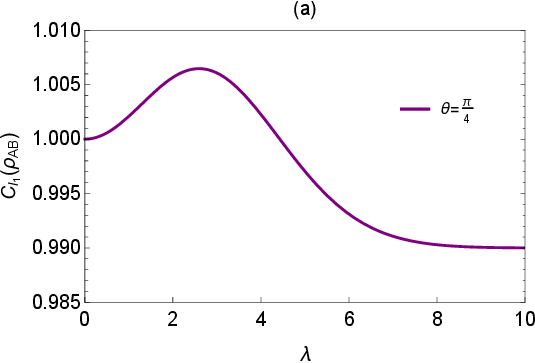}
\label{fig1a}
\end{minipage}%
\begin{minipage}[t]{0.5\linewidth}
\centering
\includegraphics[width=3.0in,height=5.2cm]{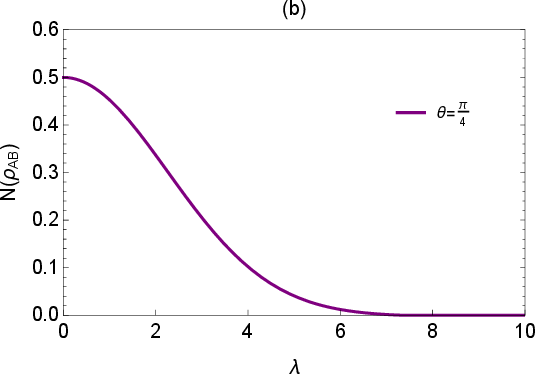}
\label{fig1b}
\end{minipage}%

\begin{minipage}[t]{0.5\linewidth}
\centering
\includegraphics[width=3.0in,height=5.2cm]{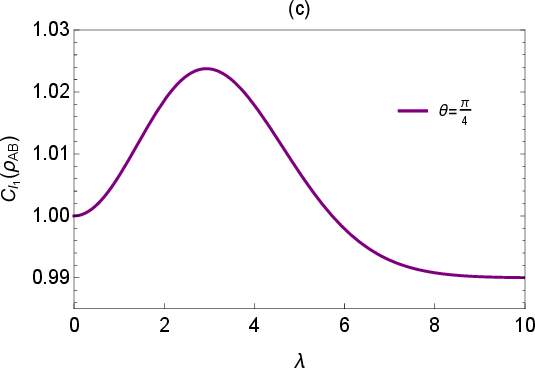}
\label{fig1c}
\end{minipage}%
\begin{minipage}[t]{0.5\linewidth}
\centering
\includegraphics[width=3.0in,height=5.2cm]{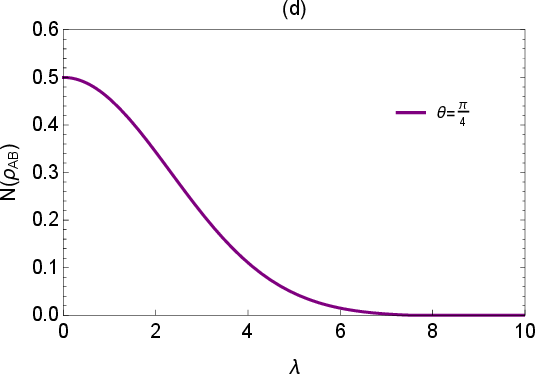}
\label{fig1d}
\end{minipage}%

\begin{minipage}[t]{0.5\linewidth}
\centering
\includegraphics[width=3.0in,height=5.2cm]{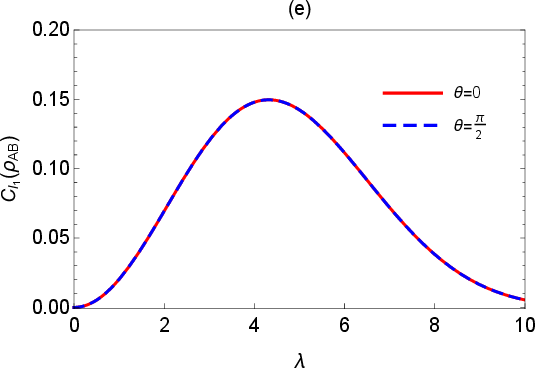}
\label{fig1c}
\end{minipage}%
\begin{minipage}[t]{0.5\linewidth}
\centering
\includegraphics[width=3.0in,height=5.2cm]{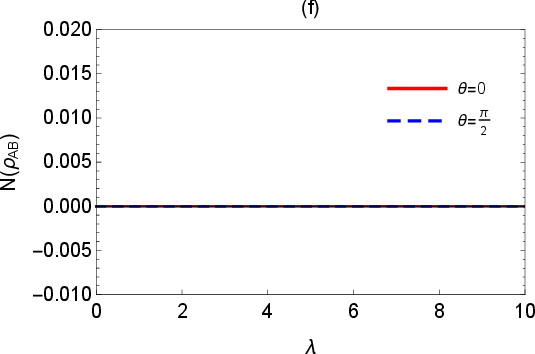}
\label{fig1d}
\end{minipage}%

\begin{minipage}[t]{0.5\linewidth}
\centering
\includegraphics[width=3.0in,height=5.2cm]{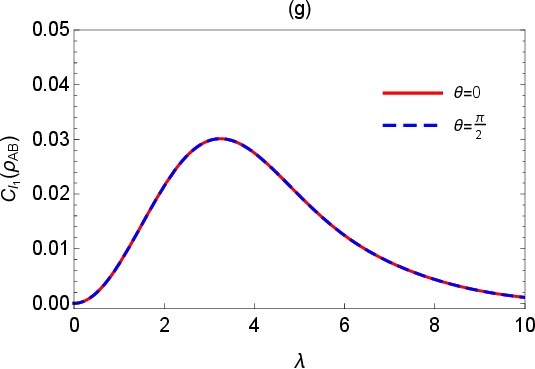}
\label{fig1c}
\end{minipage}%
\begin{minipage}[t]{0.5\linewidth}
\centering
\includegraphics[width=3.0in,height=5.2cm]{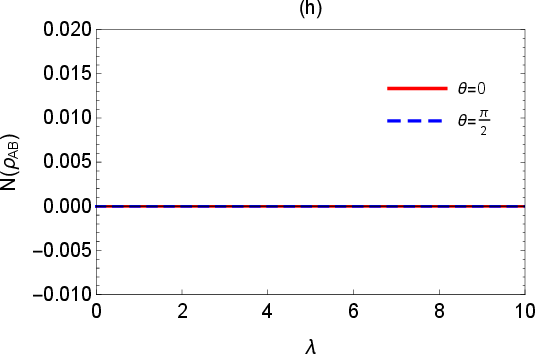}
\label{fig1d}
\end{minipage}%
\caption{ Quantum coherence and entanglement as a function of the coupling strength  $\lambda$ under two spacetime separation scenarios: panels (a), (b), (e), and (f) correspond to the lightlike case with $L/\sigma=\Delta\tau/\sigma=3$, while panels (c), (d), (g), and (h) represent the spacelike case with  $L/\sigma=5, \Delta\tau/\sigma=3$.  In all cases, the parameters are fixed as $\eta/\sigma=\Omega_{A}\sigma=\Omega_{B}\sigma=1$. The top four panels depict results for a maximally entangled initial state with \(\theta\) is $\pi/4$, whereas the bottom four panels correspond to initially separable states with $\theta=0$
or $\theta=\pi/2$. }
\label{F3}
\end{figure}

Fig.\ref{F3} illustrates the behavior of  quantum coherence \(C_{l_1}(\rho_{AB})\) and quantum entanglement \(N(\rho_{AB})\) as a function of the coupling strength \(\lambda\).  For the four subgraphs above, the detectors are initially prepared in a maximally entangled state, corresponding to \(\theta = \pi/4\). When \(\lambda = 0\),
the detectors do not interact with the field, and the system retains its initial maximal entanglement and coherence.
As the coupling strength \(\lambda\) increases, quantum coherence exhibits a non-monotonic evolution: it initially rises, reaches a peak, and then gradually decreases to an asymptotic value. This behavior indicates that quantum coherence can be extracted from the vacuum when the system-field interaction is properly tuned. Specifically, when the coherence surpasses its initial value, it reflects additional coherence harvested from the vacuum field; conversely, when the coherence drops below its initial value, a portion of the system’s coherence has been redistributed into correlations between the detectors and the field. This observation highlights the dual role of detector-field interaction: it can either enhance or diminish quantum coherence depending on the interaction strength. With appropriate control, this interaction can be exploited to amplify coherence, which may have potential applications in quantum information processing and communication. In contrast, quantum entanglement exhibits a monotonic decay with increasing
\(\lambda\), eventually vanishing at strong coupling. This divergence arises from the monogamy property of entanglement: when the system interacts with the environment (i.e., the field), entanglement between the detectors is inevitably degraded due to information leakage. Quantum coherence, however, is not constrained by monogamy and can persist-or even be enhanced-through interaction with the field, demonstrating greater robustness and longevity in open quantum systems.

We next examine the case where the detectors are initially in a separable state (the bottom four panels, corresponding to (\(\theta = 0\) or \(\pi/2\))).
In this scenario, quantum entanglement remains strictly zero for all values of the coupling strength \(\lambda\), indicating that entanglement cannot be harvested from the vacuum if the initial state is unentangled. In contrast, quantum coherence shows a markedly different dynamical behavior. Although it starts at zero when \(\lambda = 0\), it rapidly increases as \(\lambda\) grows, reaches a maximum, and then gradually declines at larger  \(\lambda\). This clearly demonstrates that, even in the absence of initial quantum correlations, detectors can still extract coherence from the vacuum via their interaction with the field. This result underscores a fundamental distinction between coherence and entanglement in field-detector systems. While entanglement harvesting is heavily dependent on the initial state, coherence proves to be more flexible and resilient. It can be dynamically generated through interaction, highlighting its significance as a more accessible and robust quantum resource for relativistic quantum information tasks.

\begin{figure}
\begin{minipage}[t]{0.5\linewidth}
\centering
\includegraphics[width=3.0in,height=5.2cm]{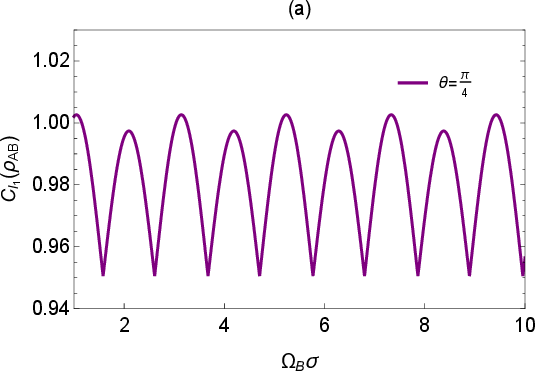}
\label{fig1a}
\end{minipage}%
\begin{minipage}[t]{0.5\linewidth}
\centering
\includegraphics[width=3.0in,height=5.2cm]{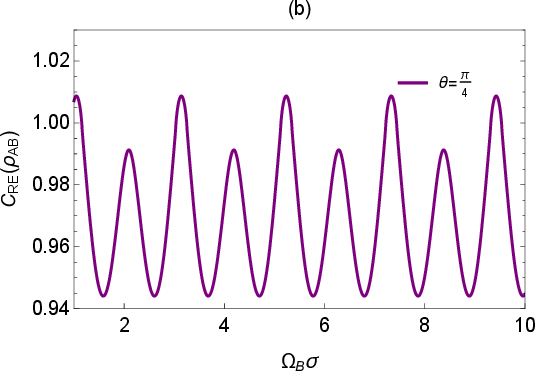}
\label{fig1b}
\end{minipage}%

\begin{minipage}[t]{0.5\linewidth}
\centering
\includegraphics[width=3.0in,height=5.2cm]{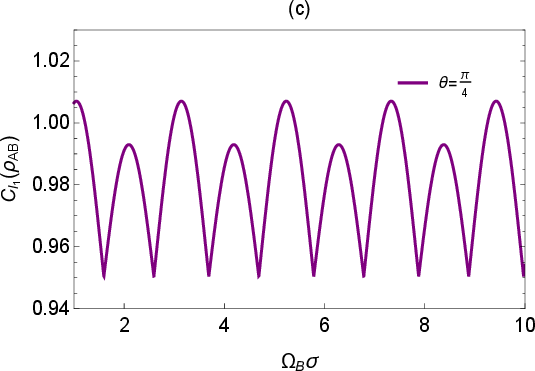}
\label{fig1c}
\end{minipage}%
\begin{minipage}[t]{0.5\linewidth}
\centering
\includegraphics[width=3.0in,height=5.2cm]{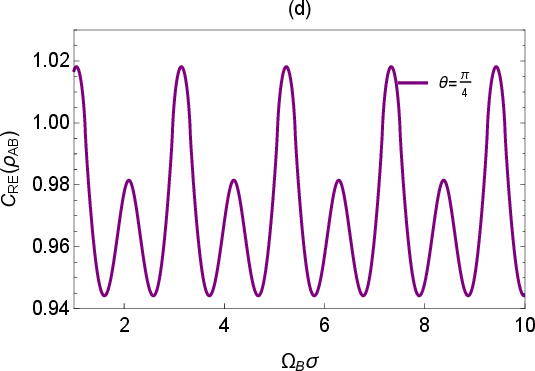}
\label{fig1d}
\end{minipage}%
\caption{Quantum coherence as a function of coupling strength $\Omega_{B}\sigma$, where the detectors are lightlike separated in panels (a) and (b) with
$L/\sigma=\Delta\tau/\sigma=3$, and spacelike separated in panels (c) and (d) with
 $L/\sigma=5, \Delta\tau/\sigma=3$. In all cases, the parameters are fixed as $\eta/\sigma=\Omega_{A}\sigma=\lambda=1, \theta=\pi/4 $.}
\label{F4}
\end{figure}

It is well known that when two detectors possess identical energy gaps, they are classified as ideal detectors; otherwise, they are referred to as non-ideal detectors. Based on this distinction, we present Fig.\ref{F4}, which investigates the influence of detector asymmetry on quantum coherence. In this Fig.\ref{F4}, the energy gap of detector $A$ is fixed at $\Omega_A\sigma=1$,
while the energy gap of detector $B$, $\Omega_B\sigma$, is treated as a tunable parameter.
This setup allows us to explore the transition from ideal  $\Omega_B\sigma=1$ to non-ideal configurations ( $\Omega_B\sigma\neq1$). The resulting plots display the behaviors of quantum coherence as functions of $\Omega_B\sigma$, under otherwise fixed system parameters. When both detectors have identical energy gaps  ($\Omega_A\sigma=\Omega_B\sigma=1$), the system is in the ideal case, and the initial values of coherence serve as benchmarks for comparison. As $\Omega_B\sigma$ deviates from $1$, periodic oscillations appear in both quantum coherence due to phase interference effect introduced by the asymmetry. Importantly, the coherence in non-ideal cases never exceeds that of the ideal configuration.
This observation highlights that ideal detectors are more efficient in harvesting quantum coherence from the vacuum. The asymmetry between detectors introduces destructive interference and phase mismatch, which degrade the system's capacity to establish or maintain coherence. Therefore, maintaining identical energy gaps is crucial for optimizing quantum coherence in field-theoretic quantum information tasks.

\section{Conclusions}
In this work, we have employed a nonperturbative framework to explore the dynamical behavior of quantum coherence between a pair of initially entangled UDW detectors interacting with a quantum scalar field in  (3+1)-dimensional Minkowski spacetime.
Specifically, we focused on the scenario where detector-field coupling is implemented via instantaneous $\delta$-function switching, which differs markedly from the more commonly studied Gaussian switching profiles.

Our findings reveal that, when the detectors are initially prepared in a maximally entangled state, the evolution of quantum coherence as a function of the coupling strength exhibits a non-monotonic pattern: it initially increases, reaching a peak value, and then gradually decays to a fixed asymptotic value. This behavior indicates a dual role played by the nonperturbative detector-field interaction. On one hand, the interaction enables the system to harvest quantum coherence from the vacuum field-reflected in coherence values that exceed the initial value. On the other hand, for stronger interactions, some of the initial coherence is redistributed into correlations between the detectors and the field, leading to a decrease in the coherence between the detectors themselves. In contrast, quantum entanglement consistently decreases monotonically with increasing coupling strength and eventually vanishes. This observation highlights that nonperturbative interactions play a purely decohering role in the context of entanglement, degrading it through information leakage to the quantum field. These results contrast with earlier studies employing perturbative or Gaussian-switching models, where both quantum coherence and entanglement were found to degrade in tandem under the influence of the Unruh effect and interaction \cite{rbm1,rbm2,rbm3}. Our nonperturbative approach thus provides new and complementary insights into the complex interplay between quantum coherence and entanglement in  quantum field theory.

Furthermore, we have extended our analysis to the case where the detectors are initially in a separable state (i.e., unentangled). In this scenario, quantum coherence-though initially zero-can still be dynamically generated via the
detector-field interaction, exhibiting a rise-fall behavior similar to the entangled case. Quantum entanglement, however, remains strictly zero across all values of the coupling strength, indicating that entanglement harvesting from the vacuum is not possible without initial entanglement in the system. This distinction underlines a fundamental asymmetry between coherence and entanglement in open quantum systems. Overall, our results emphasize the robustness and operational advantage of quantum coherence as a resource in relativistic quantum information tasks. While entanglement is fragile and highly sensitive to both initial conditions and environmental interactions, coherence is shown to be more flexible, accessible, and resilient, even in strongly interacting, nonperturbative regimes. These features suggest that quantum coherence could play a central role in the design of future quantum technologies operating in  field-theoretic settings.

\begin{acknowledgments}
This work is supported by the National Natural
Science Foundation of China (Grant Nos. 12205133) and  the Special Fund for Basic Scientific Research of Provincial Universities in Liaoning under grant NO. LS2024Q002.
\end{acknowledgments}

\appendix
\section{The elements in the density matrix}
Consider two detectors in a \((3+1)\)-dimensional Minkowski spacetime. The time-evolution operator \( \widehat{U}_{I} \) provided above Eq.(\ref{3}) reads
\begin{eqnarray}
\widehat{U}_{I}=\widehat{I}_{A}\otimes\widehat{I}_{B}\otimes\widehat{X}_{(+,+)}
+\widehat{\mu}_{A}\otimes\widehat{I}_{B}\otimes\widehat{X}_{(+,-)}+
\widehat{I}_{A}\otimes\widehat{\mu}_{B}\otimes\widehat{X}_{(-,+)}+
\widehat{\mu}_{A}\otimes\widehat{\mu}_{B}\otimes\widehat{X}_{(-,-)},
\end{eqnarray}
with
\begin{eqnarray}
\widehat{X}_{(j,k)}:=\frac{1}{4}(e^{\widehat{Y}_{B}}+je^{-\widehat{Y}_{B}})
(e^{\widehat{Y}_{A}}+ke^{-\widehat{Y}_{A}}).
\end{eqnarray}
Here, we denote $\widehat{X}_{(\pm,\pm)}=\widehat{X}_{(\pm1,\pm1)}$ \cite{E11}.

Starting from the initial state of Eq.(\ref{6}),  the final density matrix of the detectors can be described as
\begin{eqnarray}
\rho_{AB}=\cos^{2}\theta\rho_{AB}^{gg}+\cos\theta \sin\theta\rho_{AB}^{ge}
+\cos\theta \sin\theta\rho_{AB}^{eg}+\sin^{2}\theta\rho_{AB}^{ee},
\end{eqnarray}
where
\begin{eqnarray}
\rho_{AB}^{\alpha\beta}:=\textrm{Tr}_{\phi}[\widehat{U}_{I}(|\alpha_{A}\rangle\langle\beta_{A}|
\otimes|\alpha_{B}\rangle\langle\beta_{B}|\otimes|0\rangle\langle0|)\widehat{U}^{\dag}_{I}],
 \quad \alpha, \beta \in\{g,e\}.
\end{eqnarray}
For example, $\rho_{AB}^{gg}$ can be specifically written as
\begin{eqnarray}
\rho_{AB}^{gg}&=& \textrm{Tr}_{\phi}[\widehat{U}_{I}(|g_{A}\rangle\langle g_{A}|\otimes|g_{B}\rangle\langle g_{B}|\otimes|0\rangle\langle0|)\widehat{U}^{\dag}_{I}]\\&=& f_{(++++)}|gg\rangle\langle gg|+f_{(+-++)}e^{-\textrm{i}\Omega_{A}\tau_{A,0}}|gg\rangle\langle eg|+f_{(-+++)}e^{-\textrm{i}\Omega_{B}\tau_{B,0}}|gg\rangle\langle ge|\notag\\&+&f_{(--++)}e^{-\textrm{i}\Omega_{A}\tau_{A,0}}e^{-\textrm{i}\Omega_{B}\tau_{B,0}}|gg\rangle\langle ee|+f_{(+++-)}e^{\textrm{i}\Omega_{A}\tau_{A,0}}|eg\rangle\langle gg|+f_{(+-+-)}|eg\rangle\langle eg|\notag\\&+&f_{(-++-)}e^{\textrm{i}\Omega_{A}\tau_{A,0}}e^{-\textrm{i}\Omega_{B}\tau_{B,0}}|eg\rangle\langle ge|+f_{(--+-)}e^{-\textrm{i}\Omega_{B}\tau_{B,0}}|eg\rangle\langle ee|\notag\\&+&f_{(++-+)}e^{\textrm{i}\Omega_{B}\tau_{B,0}}|ge\rangle\langle gg|+f_{(+--+)}e^{-\textrm{i}\Omega_{A}\tau_{A,0}}e^{\textrm{i}\Omega_{B}\tau_{B,0}}|ge\rangle\langle eg|+f_{(-+-+)}|ge\rangle\langle ge|\notag\\&+&f_{(---+)}e^{-\textrm{i}\Omega_{A}\tau_{A,0}}|ge\rangle\langle ee|+f_{(++--)}e^{\textrm{i}\Omega_{A}\tau_{A,0}}e^{\textrm{i}\Omega_{B}\tau_{B,0}}|ee\rangle\langle gg|\notag\\&+&f_{(+---)}e^{\textrm{i}\Omega_{B}\tau_{B,0}}|ee\rangle\langle ge|+f_{(-+--)}e^{\textrm{i}\Omega_{A}\tau_{A,0}}|ee\rangle\langle ge|+f_{(----)}|ee\rangle\langle ee|,
\end{eqnarray}
with
\begin{eqnarray}
f_{(jklm)}&:=&\langle0|\widehat{X}^{\dagger}_{(j,k)}\widehat{X}_{(l,m)}|0\rangle
\\&=&(1+jl+km+jklm)+k(1+jl)\langle0|e^{2\widehat{Y}_{A}}|0\rangle+m(1+jl)\langle0|e^{-2\widehat{Y}_{A}}|0\rangle\notag\\
&+&l\langle0|e^{-\widehat{Y}_{A}}e^{-2\widehat{Y}_{B}}e^{\widehat{Y}_{A}}|0\rangle
+j\langle0|e^{-\widehat{Y}_{A}}e^{2\widehat{Y}_{B}}e^{\widehat{Y}_{A}}|0\rangle
+kl\langle0|e^{\widehat{Y}_{A}}e^{-2\widehat{Y}_{B}}e^{\widehat{Y}_{A}}|0\rangle\notag\\
&+&jk\langle0|e^{\widehat{Y}_{A}}e^{2\widehat{Y}_{B}}e^{\widehat{Y}_{A}}|0\rangle +lm\langle0|e^{-\widehat{Y}_{A}}e^{-2\widehat{Y}_{B}}e^{-\widehat{Y}_{A}}|0\rangle\notag\\
&+&jm\langle0|e^{-\widehat{Y}_{A}}e^{2\widehat{Y}_{B}}e^{-\widehat{Y}_{A}}|0\rangle
+klm\langle0|e^{\widehat{Y}_{A}}e^{-2\widehat{Y}_{B}}e^{-\widehat{Y}_{A}}|0\rangle\notag\\
&+&kjm\langle0|e^{\widehat{Y}_{A}}e^{2\widehat{Y}_{B}}e^{-\widehat{Y}_{A}}|0\rangle.
\end{eqnarray}
The expression for $f_{(jklm)}$  can be streamlined by applying the Baker-Campbell-Hausdorff (BCH) formula, which states that
\begin{eqnarray}
e^{\widehat{A}}e^{\widehat{B}}=\textrm{exp}(\widehat{A}+\widehat{B}+\frac{1}{2}[\widehat{A},\widehat{B}]
+\frac{1}{12}[\widehat{A},[\widehat{A},\widehat{B}]]-\frac{1}{12}[\widehat{B},[\widehat{A},\widehat{B}]]
+\cdot\cdot\cdot).
\end{eqnarray}
Given the commutation relation $[\widehat{Y}_{A},\widehat{Y}_{B}]=\textrm{i}\gamma$, one obtains
\begin{eqnarray}
e^{p\widehat{Y}_{A}}e^{q\widehat{Y}_{B}}e^{r\widehat{Y}_{A}}=e^{\textrm{i}\gamma q(p-r)/2}e^{(p+r)\widehat{Y}_{A}+q\widehat{Y}_B},(p,q,r\in\Re).
\end{eqnarray}
This leads to simplifications of the vacuum expectation values involved in \( f_{(jklm)} \)  \cite{E}, reducing them to expressions such as
\begin{eqnarray}
f_{j}:&=&\langle0|e^{2\widehat{Y}_{j}}|0\rangle=\textrm{exp}(-\frac{1}{2}\int d^{n}k|\alpha_{j}(\textbf{k})|^{2}),\\
f_{p}:&=&\langle0|e^{2\widehat{Y}_{B}}e^{2\widehat{Y}_{A}}|0\rangle=
\textrm{exp}[-\frac{1}{2}\int d^{n}k(|\alpha_{A}(\textbf{k})|^{2}+|\alpha_{B}(\textbf{k})|^{2}+2\alpha_{A}(\textbf{k})\alpha_{B}^{*}(\textbf{k}))]\notag\\
&=&f_{A}f_{B}e^{\omega-2\textrm{i}\gamma},
\\f_{m}:&=&\langle0|e^{2\widehat{Y}_{B}}e^{-2\widehat{Y}_{A}}|0\rangle=
\textrm{exp}[-\frac{1}{2}\int d^{n}k(|\alpha_{A}(\textbf{k})|^{2}+|\alpha_{B}(\textbf{k})|^{2}-2\alpha_{A}(\textbf{k})\alpha_{B}^{*}(\textbf{k}))]\notag\\
&=&f_{A}f_{B}e^{-\omega+2\textrm{i}\gamma},
\end{eqnarray}
where $\omega:=2\langle0|\{\widehat{Y}_{A},\widehat{Y}_{B}\}|0\rangle$ represents the vacuum expectation of the anticommutator \cite{E11}. Consequently, the quantity $f_{jklm}$ takes the form
\begin{eqnarray}
f_{jklm}&=&\frac{1}{16}((1+jl+km+jklm)+(1+jl)(k+m)f_{A}+[(l+jkm)e^{2\textrm{i}\gamma}+(j+klm)e^{-2\textrm{i}\gamma}]f_{B}\notag\\&+&[(jk+lm)e^{\omega}+(jm+kl)e^{-\omega}]f_{A}f_{B}).
\end{eqnarray}
An important observation is that $f_{(+++-)}=f_{(++-+)}=f_{(+-++)}=f_{(-+++)}=f_{(---+)}=f_{(--+-)}=f_{(-+--)}=f_{(+---)}=0$, which simplifies the structure of $\rho_{AB}^{gg}$. Adopting the computational basis  $\{|g_{A}g_{B}\rangle,|g_{A}e_{B}\rangle,|e_{A}g_{B}\rangle,|e_{A}e_{B}\rangle\}$, the reduced density matrix can be written as
\begin{eqnarray}\label{w13}
\scalebox{1}{$\rho_{AB}^{gg}=
 \left(\!\!\begin{array}{cccccccc}
\rho_{11}^{gg} & 0 & 0 & \rho_{14}^{gg}\\
0 & \rho_{22}^{gg} & \rho_{23}^{gg} & 0\\
0 & (\rho_{23}^{gg})^{*} & \rho_{33}^{gg} & 0\\
(\rho_{14}^{gg})^{*} & 0 & 0 & \rho_{44}^{gg},
\end{array}\!\!\right),
$}
\end{eqnarray}
where
\begin{eqnarray}
\rho_{11}^{gg}=f_{(++++)}, \quad \rho_{22}^{gg}=f_{(-+-+)}, \quad \rho_{33}^{gg}=f_{(+-+-)}, \quad \rho_{44}^{gg}=f_{(----)},\\
\rho_{14}^{gg}=f_{(--++)}e^{-\textrm{i}(\Omega_{A}\tau_{A,0}+\Omega_{B}\tau_{B,0})}, \quad \notag \rho_{23}^{gg}=f_{(+--+)}e^{-\textrm{i}(\Omega_{A}\tau_{A,0}-\Omega_{B}\tau_{B,0})}.
\end{eqnarray}
With $\rho_{AB}^{ge},\rho_{AB}^{eg},\rho_{AB}^{ee}$, the total density matrix is
\begin{eqnarray}\label{w13}
\scalebox{1}{$\rho_{AB}=
 \left(\!\!\begin{array}{cccccccc}
\rho_{11} & 0 & 0 & \rho_{14}\\
0 & \rho_{22} & \rho_{23} & 0\\
0 & \rho_{23}^{*} & \rho_{33} & 0\\
\rho_{14}^{*} & 0 & 0 & \rho_{44}
\end{array}\!\!\right),
$}
\end{eqnarray}
\begin{eqnarray}
\rho_{11}&=&\cos^{2}\theta f_{(++++)}+\cos\theta \sin\theta f_{(--++)}e^{\textrm{i}(\Omega_{A}\tau_{A,0}+\Omega_{B}\tau_{B,0})}\notag\\
&+&\cos\theta \sin\theta f_{(++--)}e^{-\textrm{i}(\Omega_{A}\tau_{A,0}+\Omega_{B}\tau_{B,0})}+\sin^{2}\theta f_{(----)},
\end{eqnarray}
\begin{eqnarray}
\rho_{22}&=&\cos^{2}\theta f_{(-+-+)}+\cos\theta \sin\theta f_{(+--+)}e^{\textrm{i}(\Omega_{A}\tau_{A,0}+\Omega_{B}\tau_{B,0})}\notag\\
&+&\cos\theta \sin\theta f_{(-++-)}e^{-\textrm{i}(\Omega_{A}\tau_{A,0}+\Omega_{B}\tau_{B,0})}+\sin^{2}\theta f_{(+-+-)},
\end{eqnarray}
\begin{eqnarray}
\rho_{33}&=&\cos^{2}\theta f_{(+-+-)}+\cos\theta \sin\theta f_{(-++-)}e^{\textrm{i}(\Omega_{A}\tau_{A,0}+\Omega_{B}\tau_{B,0})}\notag\\
&+&\cos\theta \sin\theta f_{(+--+)}e^{-\textrm{i}(\Omega_{A}\tau_{A,0}+\Omega_{B}\tau_{B,0})}+\sin^{2}\theta f_{(-+-+)},
\end{eqnarray}
\begin{eqnarray}
\rho_{44}&=&\cos^{2}\theta f_{(----)}+\cos\theta \sin\theta f_{(++--)}e^{\textrm{i}(\Omega_{A}\tau_{A,0}+\Omega_{B}\tau_{B,0})}\notag\\
&+&\cos\theta \sin\theta f_{(--++)}e^{-\textrm{i}(\Omega_{A}\tau_{A,0}+\Omega_{B}\tau_{B,0})}+\sin^{2}\theta f_{(++++)},
\end{eqnarray}
\begin{eqnarray}
\rho_{14}&=&\cos^{2}\theta f_{(--++)}e^{-\textrm{i}(\Omega_{A}\tau_{A,0}+\Omega_{B}\tau_{B,0})}+\cos\theta \sin\theta f_{(++++)}\notag\\
&+&\cos\theta \sin\theta f_{(----)}e^{-\textrm{i}2(\Omega_{A}\tau_{A,0}+\Omega_{B}\tau_{B,0})}+\sin^{2}\theta f_{(++--)}e^{-\textrm{i}(\Omega_{A}\tau_{A,0}+\Omega_{B}\tau_{B,0})},
\end{eqnarray}
\begin{eqnarray}
\rho_{23}&=&\cos^{2}\theta f_{(+--+)}e^{-\textrm{i}(\Omega_{A}\tau_{A,0}-\Omega_{B}\tau_{B,0})}+\cos\theta \sin\theta f_{(-+-+)}e^{\textrm{i}2\Omega_{B}\tau_{B,0}}\notag\\
&+&\cos\theta \sin\theta f_{(+-+-)}e^{-\textrm{i}2\Omega_{A}\tau_{A,0}}+\sin^{2}\theta f_{(-++-)}e^{-\textrm{i}(\Omega_{A}\tau_{A,0}-\Omega_{B}\tau_{B,0})},
\end{eqnarray}
\begin{eqnarray}
f_{(\pm\pm\pm\pm)}=\frac{1}{4}[1\pm f_{A}\pm f_{B}\cos(2\kappa)+f_{A}f_{B}\cosh\omega],
\end{eqnarray}
\begin{eqnarray}
f_{(\pm\pm\mp\mp)}=\mp\frac{1}{4}f_{B}[\textrm{i}\sin(2\kappa)\mp f_{A}\sinh\omega],
\end{eqnarray}
\begin{eqnarray}
f_{(\pm\mp\pm\mp)}=\frac{1}{4}[1\mp f_{A}\pm f_{B}\cos(2\kappa)-f_{A}f_{B}\cosh\omega],
\end{eqnarray}
\begin{eqnarray}
f_{(\pm\mp\mp\pm)}=\mp\frac{1}{4}f_{B}[\textrm{i}\sin(2\kappa)\pm f_{A}\sinh\omega].
\end{eqnarray}

The expressions above can be further simplified by introducing the shorthand \( \gamma = \Omega_{A} \tau_{A,0} + \Omega_{B} \tau_{B,0} \). For example, consider  \( \rho_{11} \). We rewrite the term \( f_{(\pm\pm\pm\pm)} = \frac{1}{4}(P \pm Q) \), where \( P := 1 + f_{A}f_{B}\cosh\omega \) and \( Q := f_{A} + f_{B}\cos(2\kappa) \).
With these definitions,  \( \rho_{11} \)  can be expressed succinctly as
\begin{eqnarray}
\rho_{11}&=&\frac{1}{4}[P+(2\cos^{2}\theta-1)Q]+\frac{f_{B}}{2}
\cos\theta \sin\theta[f_{A}\sinh\omega \cos\gamma-\sin(2\kappa)\sin\gamma].
\end{eqnarray}
Using similar rearrangements and substitutions, one obtains the full density matrix elements in a compact form.

\end{document}